# Magnetic-interaction-induced superconductivity in metals

College of Applied Sciences, Beijing University of Technology, Beijing, China, 100124*

Jiang Jinhuan

**Abstract:** In this paper, a microscopic theory of magnetic-interaction-induced pairing in superconductivity of metals was developed on the basis of four idealized assumptions: (1) only a small number of electrons are involved in superconductivity; (2) magnetic interactions between electron spins lead to superconductivity; (3) there are different electronic states, i.e., doubly-occupied, singly-occupied (spin up or down) and empty states; (4) the average kinetic energy of electrons complies with the equipartition theorem of energy. A formula to estimate $T_C$ was thus derived. It was found that, $T_C$ is not only related to the electron density and the critical magnetic field, but also to the degrees of freedom of electrons. The $T_C$ values calculated from this formula are in good agreement with the experimental results for most metals. According to this theory, $T_C$ generally increases with decreasing dimension of metals. For example, $T_C$ in the 3-dimensional (3D) Al metal is 1.19K, but increases to 1.46K in 2D and 2.06K in 1D.

PACS numbers: 74.20. Mn, 75.47. Np, 74.20. –z, 74.20. De

## 1. Four idealized assumptions

In 1911, Onnes found near zero resistance in mercury at 4.2K [1]. The temperature, where a superconductor loses its resistance, is called the superconducting transition temperature $T_C$. In 1914, Onnes found that the superconducting state can be destroyed by an external magnetic field [2]. When the external magnetic field is greater than the critical magnetic field $B_C$, electrical resistance suddenly appears, and thus the superconducting state is converted to the normal state. In 1934, Gorter and Casimir first proposed a phenomenological two-fluid model to explain superconductivity [3]. The central point of this model is that there are two types of electrons in superconductors: normal electrons and superconducting electrons. In 1956, Cooper made an important step in establishing the microscopic theory of superconductivity [4], and could prove that the paired electrons can form bound states near the Fermi surface no matter how weak the effective attraction between electrons is. In 1957, Bardeen, Cooper and Schrieffer developed a microscopic theory for superconductivity which was called BCS theory [5]. According to the BCS theory, the electrons near the Fermi surface can pair up (called Cooper pairs), forming bound states via electron-phonon interactions, thus lowering the energy of the system. The Cooper pairs can undergo Bose-Einstein condensation and make the material superconducting. So far, Cooper pairs are found to exist in all kinds of superconductors, and the breakdown of Cooper pairs is the main reason to destroy superconductivity [6]. The energy needed to break up a Cooper pair, i.e., the bonding energy, is the double of the BCS superconducting energy gap $\Delta$. This energy gap is an important energy scale in superconductors that can determine $T_C$. For an ideally isotropic s-wave superconductor, it was found that the energy gap is given by $\Delta = 3.52 k_B T_C/2$. The microscopic BCS theory of superconductivity is successful to describe the conventional superconductors, but it cannot explain the high-Tc superconductors because the highest $T_C$ given by the McMillan limit is 40K. Therefore, we proposed a magnetic-interaction-induced pairing mechanism for metals based on the following four idealized assumptions.

(1) Only a small number of electrons near the Fermi surface contribute to superconductivity. The electron density $n$ is given by the sum of the normal electrons $n_n$ and the superconducting electrons $n_S$, i. e., $n = n_n + n_S$. The density of superconducting electrons is $T_C/T_F$ times the electron density (at T = 0K), i.e., $n_S = (T_C/T_F)n$, where $T_F = \hbar^2 (3\pi^2 n)^{2/3}/2m_e k_B$ is the Fermi temperature.

---

*E-mail: jiangjh@bjut.edu.cn       1

(2) Magnetic interactions between electron spins lead to the formation of electron Cooper pairs near the Fermi surface, and give rise to superconductivity; and all other kinds of interactions can be neglected. This magnetic interaction energy is exactly the condensed energy which is given by the sum of the bonding energy of the Cooper pairs, i.e., $n_S \Delta$. An external magnetic field can destroy this magnetic interaction between the superconducting electrons, break up Cooper pairs, and transform the superconducting states into normal states. Therefore, the maximum value of the condensed energy density equals the maximum energy density of the external magnetic field $B_C^2/2\mu_0$.

(3) There are different electronic states in metals, i.e., doubly-occupied states, singly-occupied (spin up or down) states, and empty states. The electron doubly-occupied states form the superconducting states (or Cooper pairs) while the singly-occupied states form the normal states.

(4) The average kinetic energy of electrons complies with the equipartition theorem of energy. Based on the above theorem, the maximum average kinetic energy of the superconducting electrons is $ik_B T_C/2$ ($i = 1,2,3,\cdots$), where $i$ the electronic degrees of freedom. An increase in temperature can break up Cooper pairs, and thus destroy superconductivity. Therefore, the bonding energy of a Cooper pair is equal to $2\Delta = ik_B T_C$, and thus the superconducting energy gap is given by $\Delta = ik_B T_C/2$. For the ideally isotopic s-wave superconductors in metals, the superconducting energy gap is $\Delta = 3k_B T_C/2$, which is close to the experimental results of $\Delta = 3.52 k_B T_C/2$.

## 2. The superconducting transition temperature

As stated before, only a small number of electrons contribute to superconductivity in metals, and the number of superconducting electrons reaches to its maximum at T = 0K, but still far less than the total number of free electrons i.e., $n_{S,max} \ll n$. The maximum density of the superconducting electrons is

$$n_{S,max} = (T_C/T_F)n \neq n. \tag{1}$$

These superconducting electrons can generate Cooper pairs via magnetic interactions, and form bound states, which reduces the energy of the system and for sufficiently low temperatures turns metals into superconductors. With increasing temperature the magnetic interaction in Cooper pairs will be destroyed, leading to a decrease of the number of superconducting electrons. At $T_C$, all Cooper pairs are broken up, and the number of superconducting electrons drops to zero. The energy density in a superconducting system is given by the sum of the electronic potential and kinetic energy

$$E = -n_S(T)\Delta + n_S(T)\frac{i}{2}k_B T. \tag{2}$$

If $E \leq 0$, metals will be in the superconducting states, and thus we get $T \leq (2\Delta/ik_B) = T_C$. This is exactly the critical temperature $T_C$, and $\Delta = ik_B T_C/2$.

An external magnetic field can also destroy Cooper pairs, leading to a decrease of the number of superconducting electrons. When a critical magnetic field $B_C$ is reached, all the Cooper pairs are broken up, and the number of superconducting electrons becomes zero. The energy density in a superconducting system under an external magnetic field is given by

$$E = -n_S(T)\Delta + \frac{B^2(T)}{2\mu_0} + n_S(T)\frac{i}{2}k_B T. \tag{3}$$

If $E \leq 0$, metals will be in the superconducting states, and at T = 0K we get $B \leq \sqrt{2\mu_0 n_S(0)\Delta}= B_C$. This is exactly the critical magnetic field $B_C$. Thus, we obtain

$$n_{S,max}\frac{i}{2}k_B T_C = \frac{B_C^2}{2\mu_0}. \tag{4}$$

That is, the maximum condensed energy density in metals equals the maximum energy density of the external magnetic field. Inserting Eq. (1) into Eq. (4), we obtain for the critical temperature

$$T_C = \frac{\hbar(3\pi^2)^{1/3}}{k_B\sqrt{2m_e\mu_0}} \times \frac{B_C}{n^{1/6}\sqrt{i}} = 1.56 \times 10^7 \times \frac{B_C}{n^{1/6}\sqrt{i}}. \tag{5}$$

Where $i$ is the degrees of freedom of electrons, $k_B$ is the Boltzmann constant, $m_e$ is the electron mass, $\hbar$ is the reduced Planck constant. From this formula we can see that, $T_C$ is not only related to the critical magnetic field $B_C$ and the electron density $n$, but also related to the degrees of freedom of electrons $i$.

## 3. Comparison between theory and experiment

The $T_C$ values for 21-kinds of metals calculated for different degrees of freedom $i$ are in good agreement with the experimental results as shown in the following table. It is still an open question whether and why it can be these values.

Comparison between theoretical and experimental $T_C$ of 21-kinds of metals [7]

| No. | Degrees of freedom $i$ | Metals | Electron density $n(10^{29}/m^3)$ | Critical magnetic fiel $B_C(10^{-4}T)$ | Theory $T_C$(K) | Experiments $T_C$(K) |
|---|---|---|---|---|---|---|
| 1 | $i = 3$ | Al | 1.812 | 99 | 1.19 | 1.19 |
| 2 | | Sn | 1.482 | 303 | 3.76 | 3.72 |
| 3 | | In | 1.145 | 279 | 3.62 | 3.41 |
| 4 | | W | 3.803 | 1.15 | 0.012 | 0.012 |
| 5 | | Mo | 3.840 | 93 | 0.98 | 0.92 |
| 6 | | Tl | 1.048 | 180 | 2.37 | 2.38 |
| 7 | | Os | 5.692 | 65 | 0.64 | 0.65 |
| 8 | | Hg($\alpha$) | 4.878 | 409 | 4.16 | 4.15 |
| 9 | | Ir | 6.320 | 15 | 0.146 | 0.14 |
| 10 | $i = 5$ | Re | 4.646 | 205 | 1.63 | 1.70 |
| 11 | | Ru | 5.859 | 66 | 0.505 | 0.49 |
| 12 | | Ti($\alpha$) | 2.284 | 56 | 0.50 | 0.49 |
| 13 | $i = 6$ | Pb | 1.322 | 804 | 7.19 | 7.20 |
| 14 | $i = 7$ | Th($\alpha$) | 1.214 | 163 | 1.37 | 1.37 |
| 15 | $i = 2$ | Zn | 1.315 | 54 | 0.84 | 0.84 |
| 16 | | Zr | 1.721 | 47 | 0.70 | 0.73 |
| 17 | | Cd | 0.9624 | 30 | 0.49 | 0.52 |
| 18 | $i = 1$ | Ga | 1.529 | 55 | 1.18 | 1.1 |
| 19 | $i = 12$ | Ta | 2.777 | 823 | 4.6 | 4.48 |
| 20 | | V | 3.522 | 1020 | 5.48 | 5.45 |
| 21 | $i = 16$ | Nb | 2.776 | 1950 | 9.44 | 9.26 |

According to BCS theory, the electron-phonon interaction leads to Cooper pairs, and the maximum $T_C$ estimated by the BCS theory is less than 40K. Based on a magnetic-interaction mechanism as derived Eq. (5), $T_C$ becomes higher when the critical magnetic field $B_C$ is larger or the degrees of the freedom $i$ is smaller. Therefore, we estimated that the maximum of $T_C$ is 37.8K in case of $i = 1$, $B_C = 0.1950T$ and $n = 2.776 \times 10^{29}/m^3$. We also estimated that $T_C$ is 1.46K for 2-dimensional Al and 2.06K for 1-dimensional Al. Therefore, $T_C$ is higher in materials with lower dimensions.

## 4. Conclusions

In this paper, we proposed a microscopic theory of magnetic-interaction-induced pairing in superconductivity of metals on the basis of four idealized assumptions. A formula of $T_C$ was derived. It was found that, $T_C$ is not only related to the electron density $n$ and the critical magnetic field $B_C$, but also to the degrees of freedom of electrons $i$. The $T_C$ values calculated from this formula are in good agreement with the results obtained from experiments for most metals. According to this theory, a lower dimension generally leads to a higher $T_C$. For example, $T_C$ in the two-dimensional Al metal is 1.46K, but increases to 2.06K in the one-dimensional case. The magnetic-interaction-induced mechanism of high-Tc superconductors is in progress.

## References

[1] H. K. Onnes, Commun Phys. Lab. University Leiden, 1911, No. 120b, 122b.

[2] H. K. Onnes, Commun Phys. Lab. University Leiden, 1914, No. 139f.

[3] C. J. Gorter, H. Casimir, On superconductivity I., Physica, 1934, 1: 306-320.

[4] L. N. Cooper, Bound Electron Pairs in a Degenerate Fermi Gas Theory of Superconductivity, Phys. Rev., 1956, 104: 1189-1189.

[5] J. Bardeen, L. N. Cooper, J. R. Schrieffer, Theory of Superconductivity, Phys. Rev., 1957, 108: 1175-1203.

[6] T. Xiang, d-wave Superconductors, 2007, p17-p18, Beijing: Science Press.

[7] Y. H. Zhang, Superconductor Physics, 2009, p8-p11, Hefei: University of Science and Technology of China Press; X. Z. Wu, Z. X. Li, S. P. Chen translation (Germany, Horst StÖcker), Handbook of Physics, 2004, p982-p983, Beijing: Peking University Press.